\documentclass[11pt,twoside]{article}
\usepackage{asp2010_JCB}
\usepackage{graphicx}

\resetcounters

\begin{document}

\addtocounter{page}{228}

\title{ Constraints on the Galactic Magnetic field from the Canadian Galactic Plane Survey}
\author{Kyle Rae and Jo-Anne Brown}
\affil{Department of Physics and Astronomy, University of Calgary, Canada}

\begin{abstract}
The Galactic magnetic field is important in the dynamics of our Galaxy. It is believed to play a role in star formation and influence the structure of the Galaxy. In order to understand how the Galactic magnetic field originally formed or how it is evolving, we must first determine its present topology. To this end, we have used observations from the Canadian Galactic Plane Survey (CGPS) to calculate the highest source density of rotation measures (RM) to date in the disk of the Galaxy.  Using these data, we estimate the  Galactic longitude of the RM null point in the outer Galaxy (where the RMs of extragalactic sources are observed to pass through zero, on average,  with increasing Galactic longitude).  We have also examined the RM scale height using the CGPS latitude extension.  The values of these parameters offer critical constraints for modeling the large-scale magnetic field in the Galactic disk. 
\end{abstract}

\vspace{-8mm}
\section{Introduction}

Despite significant advances in the understanding of the Galaxy's magnetic field, the large scale topology is still not well understood.  Part of the difficulty in determining the structure of the Galactic magnetic field is due to the vast distances involved. With current propulsion methods it would take about 80,000 years for a small probe to reach the nearest star \citep{spacecraft}. Therefore, direct measurements of the Galaxy's magnetic field with magnetometers would take an unacceptably long time. Furthermore, magnetic fields themselves do not radiate.  Consequently,  indirect observational methods must be used.

Most of the information we  have about the Galactic magnetic field has been derived from radio observations of polarised sources as follows. 
Since a significant fraction of the interstellar medium consists of a plasma with a frozen-in magnetic field \citep{katia},  it acts as a birefringent material to radio waves that pass through it.  As a result,  the polarisation angle of a  wave will  rotate as it propagates through the medium. This effect, known as Faraday rotation, is quantified using the aptly named Rotation Measure (RM), defined as
\begin{equation}
\label{rmequation}
{\rm{RM}} = 0.812 \int n_e \; {\bf{B}} \cdot {\rm{d{\bf{l}}}},
\end{equation}
where $n_e$ is the electron density, ${\bf{B}}$ is the magnetic field, and d{\bf{l}} is the pathlength element, which is always defined to be from the source to the receiver. 
The amount of rotation a radio wave typically experiences is linearly dependent on the square of the wavelength ($\lambda$), with the slope of the dependence being the RM such that
\begin{equation}
\label{pangle}
\tau = \tau_\circ + \lambda^2 {\rm{RM}}
\end{equation}
where $\tau$ is the observed polarisation angle, and $\tau_\circ$ is the emitted polarisation angle.
Assuming all emissions from a given source are at the same $\tau_\circ$, measuring a source of polarised radiation at multiple wavelengths will allow us to easily determine the RM for a given source.    If we know something about the distribution of free electrons along a given line of sight, we can then work backwards to determine what the magnetic field might be to create the RMs we observe. 
Using RMs  from sources both inside (puslars) and outside (extragalactic sources or EGS) the Galaxy allows us to 
further constrain the possible 
topology of the field.  The more sources we have, the more accurate the inferred magnetic field is likely to be.

\section{Observations}

The Canadian Galactic Plane Survey \citep[CGPS;][]{Taylor03} represents the most extensive Galactic disk survey done to date.  It successfully surveyed all of the second Galactic quadrant, a significant fraction of the first quadrant, and part of the third quadrant, at multiple wavelengths and in full polarisation.   The data collected by the Synthesis telescope at the Dominion Radio Astrophysical Observatory was for  four, 7.5 MHz bands, spanning a 35 MHz window centered on the 1420 MHz spectral line.  The first RMs published from these data totaled 380 EGS sources \citep{btj03}. 
These data were used to show conclusively that there are no magnetic field reversals{\footnote{Magnetic field reversal: a region of magnetic shear characterized by a roughly $180^\circ$ change in the direction of the magnetic field, typically with Galactic radius.}} in the outer Galaxy \citep{bt01, Brown03}.  

\begin{figure}[h] 
\begin{centering}
\includegraphics[width=0.9\hsize]{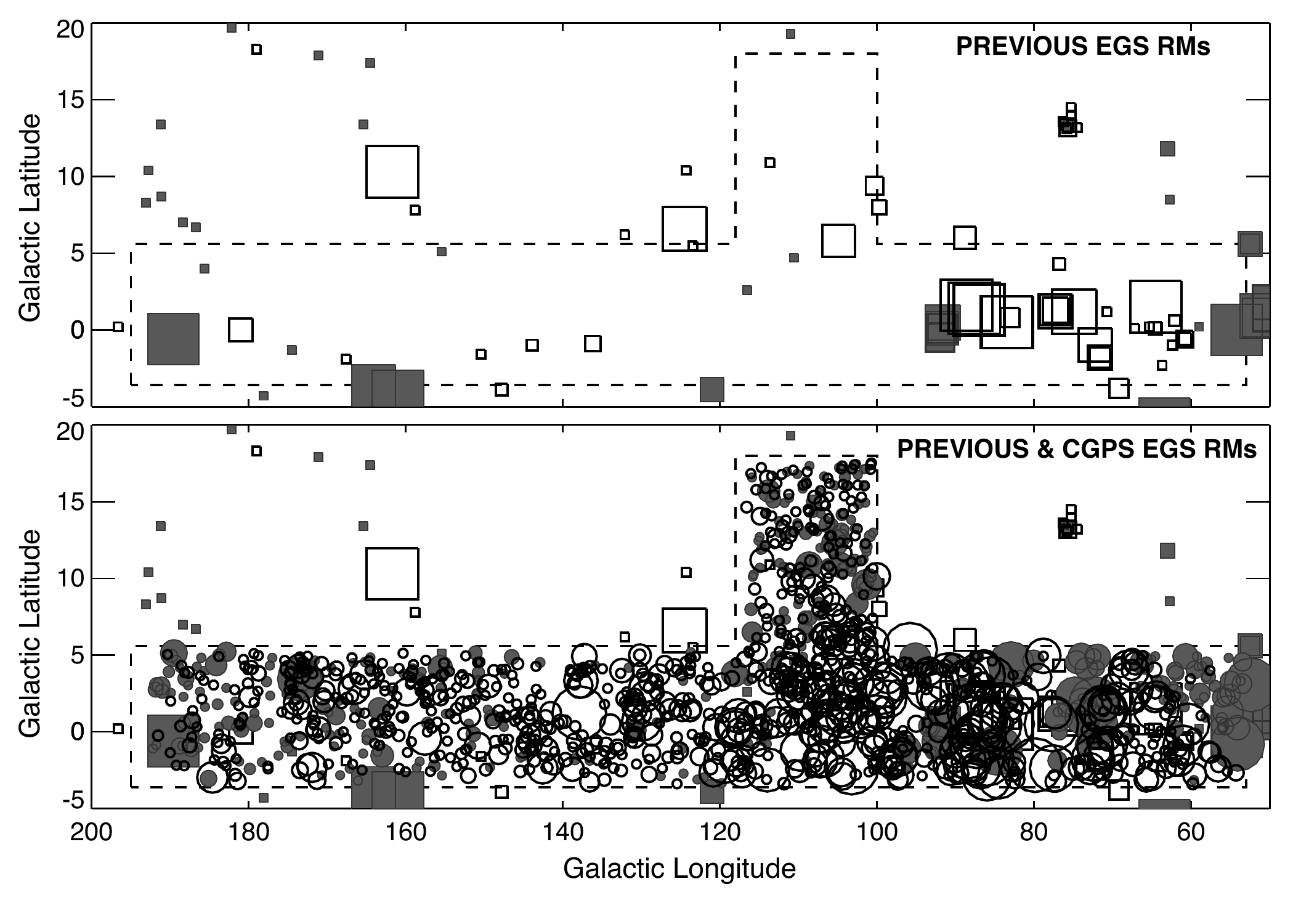}  
\caption{Rotation measures of extragalactic sources in the CGPS. Top panel: previously observed extragalactic source (EGS) rotation measures displayed as squares; Bottom panel: CGPS (circles) and EGS (squares) rotation measures. Gray filled symbols indicate positive rotation
measures, and black open symbols indicate negative rotation measures; sizes of symbols are linearly proportional to the magnitude of RM truncated between 100 and 600 rad m$^{-2}$, so that sources with $|$RM$|<100$ rad m$^{-2}$ are set to 100 rad m$^{-2}$, and those with $|$RM$|> 600$ rad m$^{-2}$ are set to 600 rad m$^{-2}$.  The dashed line delineates the CGPS region. The source density of the CGPS RMs is roughly 1 source per square degree.}
\label{rmplot}
\end{centering}
\end{figure}

Since these original publications, we have completed the observations,  data processing, and RM calculations. 
Using a list of compact source candidates identified by A.R. Taylor (private communication), we were able to 
determine reliable RMs for 1316 sources spanning the entire CGPS, using the method outlined in \citet{btj03}.  These data are illustrated in Figure \ref{rmplot}. 

\section{Topology of the Galactic Magnetic Field}

 The Galactic magnetic field is generally divided in to two distinct components: the small-scale field with scale lengths on the order of 10 parsecs and the large-scale field with scale sizes on the order of kiloparsecs. Our research focuses on the large-scale field in the disk of the Galaxy, particularly the pitch angle of the  field in the outer Galaxy, and how the field transitions from the disk to the halo. 

The high source density of the CGPS RM data  (1 source per square degree) permits statistical analyses of the RMs to study large-scale features of the field,  which otherwise might be masked by  fluctuations in the RMs caused by either the small-scale field, or by structures from within sources themselves.    As a result, we are in a position to study  features of the magnetic field that were previously inaccessible with the older data.   Here, we discuss two critical observations of the field:  the RM null point, and the RM scale height.

\subsection{The RM Null Point and the Pitch Angle of the Magnetic Field}

It is often assumed the Galactic magnetic field follows the spiral arms everywhere within the Galaxy \citep[e.g.][]{wc04,HanSpiral}. 
However, some studies have suggested the field has a much smaller pitch angle than the optical spiral arms \citep{vallee08}, or that at least the reversals  within the field  may not be aligned with the arms \citep{brown07, Sun08}.   In order to address the question of how the field is aligned relative to the optical spiral arms, 
we have used the CGPS RM data to determine the
 location of a key Galactic longitude, which we call the RM null point,  where EGS RMs are zero, on average.  Identifying the location of the RM null point  should help elucidate the relationship between the magnetic field
 and the optical spiral arms in the outer Galaxy. 

As illustrated in Figure 1, the RMs of the CGPS are 
 consistently negative in the central longitudes of the CGPS, with more positive RMs observed towards the edges of the survey.  
In the outer Galaxy, lines of sight through the disk are essentially always looking {\it{across}} spiral arms, in contrast to 
the inner Galaxy, where lines-of-sight can have a significant fraction of the path be along spirals arms.  As a result,
the trend in the RM dependence with Galactic longitude will be strongly influenced by how the field is aligned relative
to the spiral arms.  
The longitude where the RMs transition from being predominantly negative to predominantly positive  in the outer Galaxy is what we are after in this study. 

 RMs are an integrated effect over the line of sight, as demonstrated in Equation 1.   Therefore, in order for the
 observed RMs go to zero in a non-trivial way, the average $n_e$-weighted {\bf{B}} must be zero over the line-of-sight.
  There are two straightforward ways in which this could happen.  The first is to have at least one magnetic field reversal along the line of sight so that  integrated effect on the RM before and after the reversal exactly balance.  This type of `RM-zeroing' has been
  observed with lines-of-sight towards the inner Galaxy where at least one large-scale field reversal is known to exist \citep[e.g.][]{brown07}.  This is  likely the cause of the trend in the RMs at the lower longitudes of the CGPS.    Since previous studies 
  have shown there is no strong evidence for a magnetic field reversal in the outer Galaxy, 
  the other possible explanation for the observed EGS RMs passing through zero  is to have the average line-of-sight magnetic field be perpendicular to the path.  
  If, for example, the field were purely azimuthal and had no small-scale component, then we would expect all EGS RMs to be zero at $\ell = 180^\circ$.  On the other hand, if the field were aligned with the spirals arms, we would expect the EGS RMs to be zero around 
 $\ell \sim 170^\circ$, consistent with a pitch angle on the order of $10^\circ$.  We would not expect the EGS RMs to be zero at 
 $\ell > 180^\circ$, since this would imply a spiral field with the opposite `handedness' to the observed pitch angle of the spiral arms.

\begin{figure}[h] 
\begin{centering}
\includegraphics[width=0.3\hsize, angle=-90]{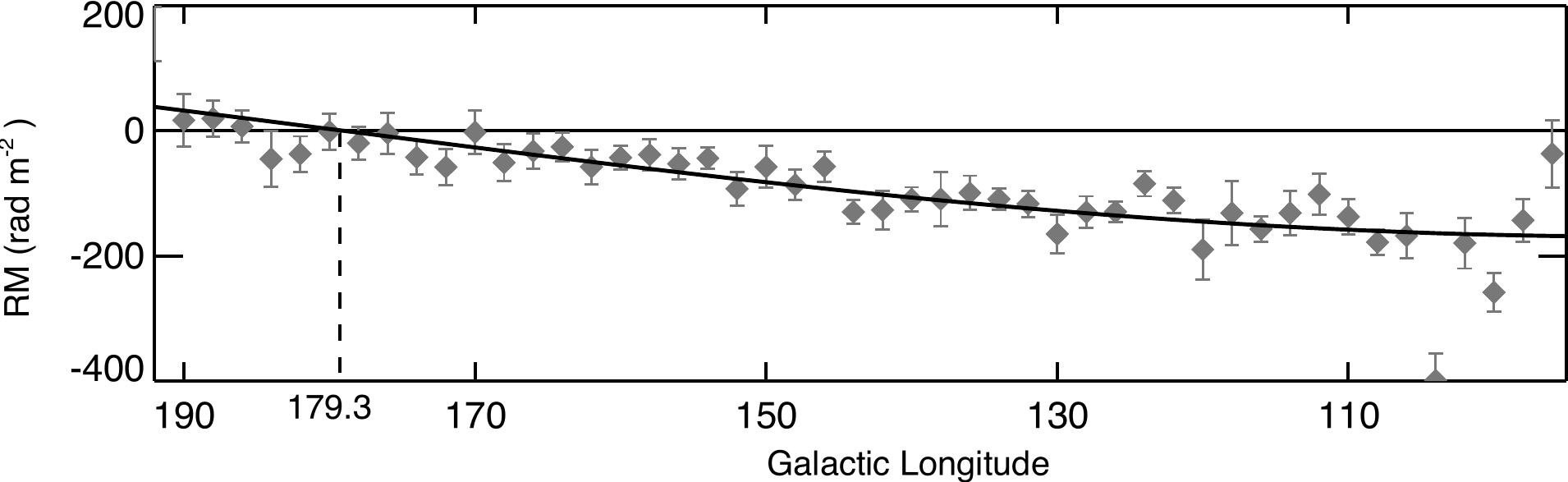} 
\caption{Determination of the RM null point in the outer Galaxy. CGPS sources between 95$^\circ$ and 195$^\circ$ longitude, averaged into 2$^\circ$ longitude bins, with 2$^\circ$ step. Each bin has a minimum of 7 sources. The error bars are the standard deviation of the mean for the bin.  The curve is a sine function fit to the data. The dotted line at 179.3$^\circ$ shows the location of the RM null point, as determined from the sine function.}
\label{rmnull}
\end{centering}
\end{figure}

In order to determine the  longitude of the RM null point, , we boxcar-averaged the CGPS RM data with $-3.5^\circ < b < 5.5^\circ$ into two degree longitude bins , with two degree steps (i.e. independent bins) between $ 95^\circ < \ell < 195^\circ$, as shown in Figure \ref{rmnull}.   Each bin has at least 7 sources. 
We then fit a sine curve to these binned data such that
$
{\rm{RM}} = {\rm{RM_{max}}} \sin(\ell_\circ - \ell),
$
and found the best-fit parameters to be $\ell_\circ = 179.3^\circ \pm 0.6^\circ$, RM$_{\rm{max}} = -174 \pm 2$ rad m$^{-2}$.    The value of
 $\ell_\circ$ suggests a large-scale field with a very small (close to zero) pitch angle, where the variations in RMs around this longitude are likely the result of localized features, such as 
supernova remnants, containing strong intrinsic magnetic fields which dominate the  line-of-sight component of the large-scale field.

\subsection{Rotation Measure Scale Height}

\begin{figure}[hb] 
\begin{centering}
\includegraphics[width=0.7\hsize]{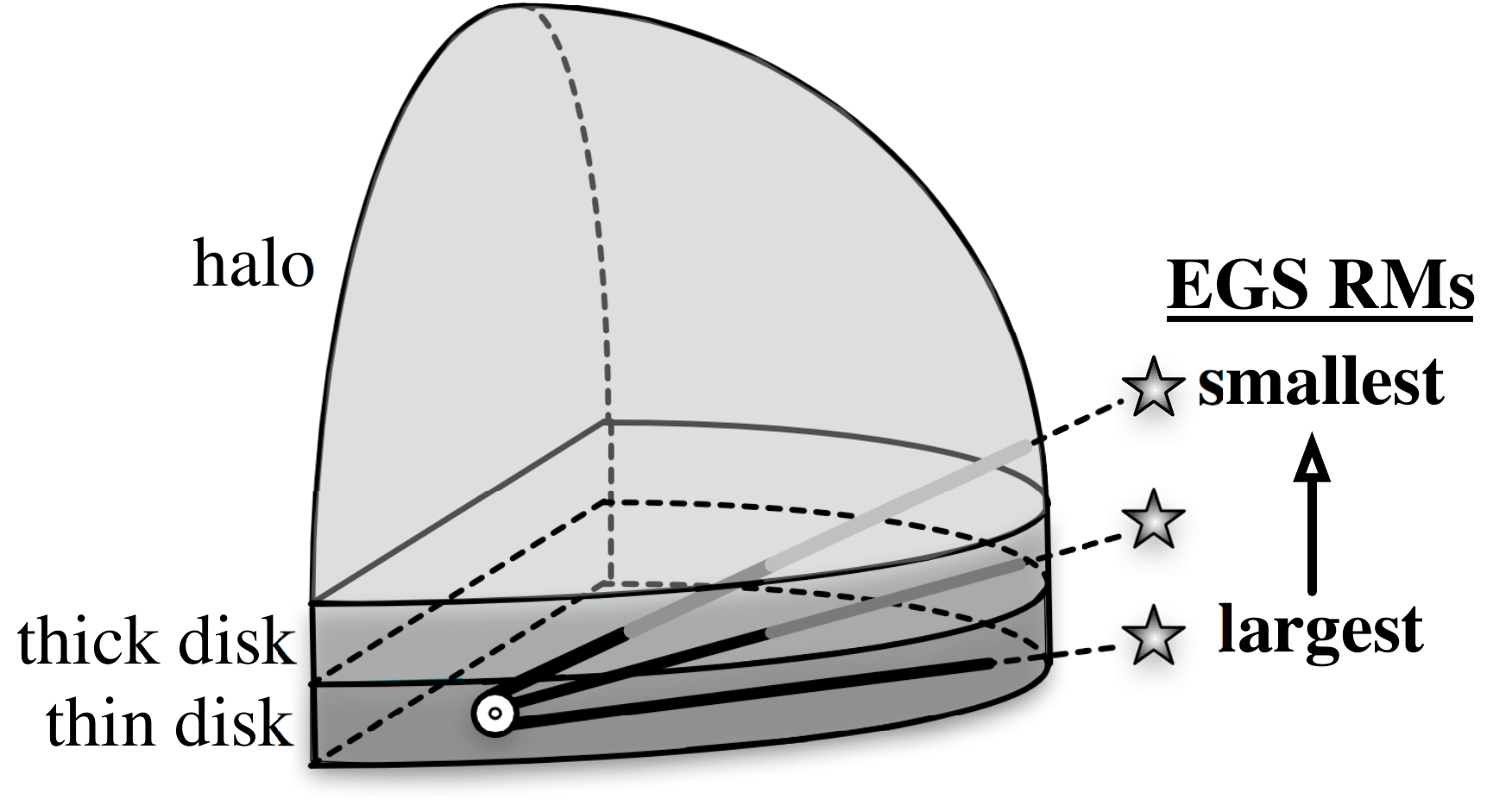}  
\caption{Cross-section (not to scale) of the Galaxy showing path length contributions from the different layers to the rotation measures. As the latitude of the observations increases, less of the total path is through the disk,  resulting in smaller rotation measures. }
\label{crosssection}
\end{centering}
\end{figure}

As illustrated in Figure \ref{crosssection},  the cross-section of the Galaxy can be described  as having three distinct  regions: the thin disk, the thick disk, and the halo.  The magnetic field is concentrated in the disk of the Galaxy, with the magnetic field in the halo 
being at least an order of magnitude weaker than that in the disk \citep{han94}.  How the magnetic field transitions between the disk and the halo has only been superficially explored.  \citet{sk80} first determined the scale height of the magnetoionic disk  as  $\sim$1.4 kpc.    While recent studies  have the scale height of the warm ionized medium estimated as 1.5 kpc \citep{gaensler08}, 
determining the scale-height of the magnetic field itself
requires an understanding of how the magnetic field is coupled to the ionized medium, which remains an open question.  Therefore, the only question that can be addressed at this stage, with the data presented in this paper,  is an assessment  of  the scale height of the RMs, also interpreted as the scale height of the magnetoionic disk for a small segment of the outer Galaxy.

\begin{figure}[h]
\begin{centering}
\includegraphics[width=0.8\hsize, angle=0]{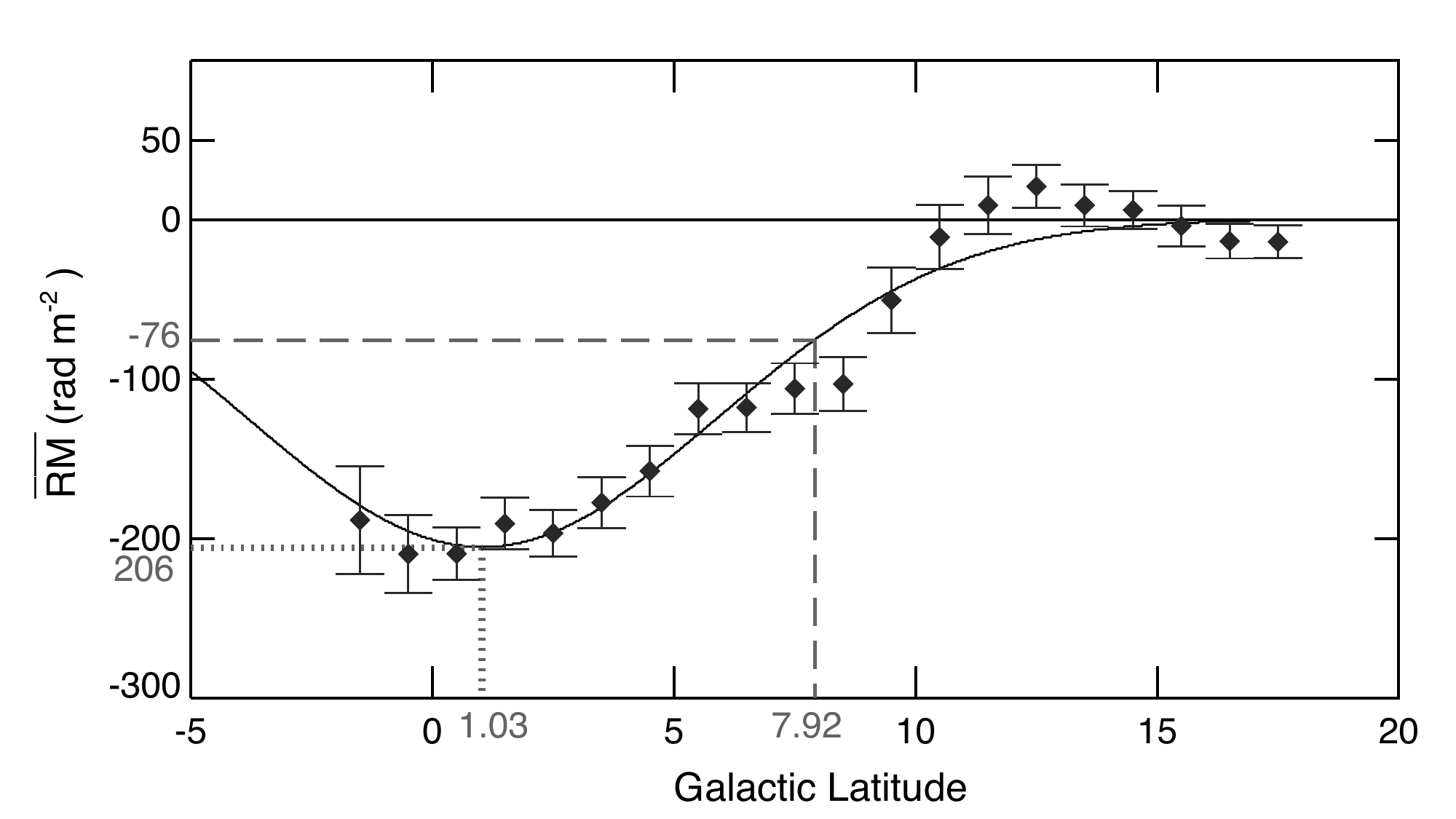}   
\caption{Rotation measure scale height of the Galaxy. CGPS rotation measures in the latitude extension  ($-3.5^\circ < b < 18^\circ$;   $100^\circ < \ell < 118^\circ$) longitude. The individual RM data have been smoothed by box-car averaging with 2$^\circ$ bins in latitude, and 1$^\circ$ step. The error bars shown are the standard deviation of the mean within the bin.  The curve is a Gaussian fit to the data, with the peak of the Gaussian indicated by the dashed lines, and  $1/e$ of the peak, representing the scale height of the RMs,   indicated by the dashed lines.  }
\label{vertplot}
\end{centering}
\end{figure}

For this analysis, we used the CGPS latitude extension data which was observed between $100^\circ < \ell < 118^\circ$, and 
$-3.5^\circ < b < 18^\circ$.  As illustrated in Figure \ref{vertplot}, we boxcar-averaged and smoothed these data, using two
degree bins, with one degree step.  Next, we fit the data with  a Gaussian curve of the form,
\begin{equation}
{\rm{RM}} = {\rm{RM}}_\circ \exp(x^2/2)  \;\;{\rm{with}} \;\; x=\frac{b - b_\circ}{\sigma},
\end{equation}
where RM$_\circ$ is the peak of the Gaussian, $b_\circ$ is the location in latitude of RM$_\circ$ and $\sigma$ is the 
width (standard deviation) of the Gaussian.  The best fit values of these parameters to the data are RM$_\circ = -206 \pm 8$ rad 
m$^{-2}$, $b_\circ = 1.03^\circ \pm 0.55^\circ$, and  $\sigma = 4.86^\circ \pm 0.45^\circ$.

The fact that the peak of the RMs is shifted from $b=0$ is consistent with the understanding that the Galactic disk is warped \citep[e.g.][]{warp}. Using the standard definition of `scale height' to refer to the  height at which the amplitude of the binned RMs goes to $1/e$ of its peak value, we find the binned RMs fall to  $\sim76$ rad m$^{-2}$, at a latitude of $b = 7.92^\circ$, as indicated on Figure 4.  

As a `back of the envelope' calculation, we can take the Galactic disk as a slab with  constant height and effective radius of 15 kpc, and use $\ell = 109^\circ$ as the nominal
longitude within the region of study.  We find that for $\Delta b = 7.92^\circ - 1.03^\circ = 6.89^\circ$, the corresponding scale height is 1.2 kpc, consistent with the findings of \citet{sk80}.

However, there are many points not considered in this calculation that require further investigation.  First, it is likely the scale height of the RMs in this direction will be dominated by the Perseus arm.   If we estimate the distance to the Perseus arm as 3.5 kpc, the scale height estimate is then reduced drastically to $\sim 0.4$ kpc.  Consequently, what needs to be determined is how the line-of-sight beyond the Perseus arm contributes to the RM. For example, the warp of the Galaxy may need to be considered further in that   lines-of-sight for some sources may in fact pass through the disk {\it{twice}} as the Galaxy warps up.   

These considerations form have led us to pursue additional observations with the Synthesis telescope south of the Galactic disk, at  essentially the same longitudes.  
These observations, which we call the Southern Latitude Extension (SLE), will use the same field spacing as the CGPS, and should result in a data set matching the CGPS in sensitivity.  These new data will allow us to clarify the RM structure  with latitude and properly analyze the scale height at this longitude.

\section{Summary}

We have determined reliable rotation measures for 1316 extragalactic sources, calculated from the four-band polarisation observations carried out by the Synthesis telescope at DRAO as part of the Canadian Galactic Plane Survey (CGPS).  These data represent more than a factor of 20 increase in the number of sources in the same region published prior to the introduction of the CGPS.  Using these data, we have shown that the magnetic field in the outer disk of the Galaxy has a very small (almost zero) pitch angle, strongly suggesting that the field is not aligned with the spiral arms in this region.  In addition, we have  estimated the rotation measure scale height in the longitude region of the CGPS latitude extension to be on the order of 1.2 kpc.  This value, however, requires more investigation, as the dominant contribution to the RMs (namely the Perseus arm) has not been considered separately in this calculation.  It is expected that new observations south of the Galactic disk, which are currently underway, will contribute considerably to the understanding of the disk-halo transition for the magnetic field in this region.

\acknowledgements 

The DRAO Canadian Galactic Plane Survey is a Canadian
project with international partners. The Dominion Radio Astrophysical
Observatory is operated as a national facility by the National Research
Council of Canada. The Survey is supported by a grant from the Natural
Sciences and Engineering Research Council of Canada. 
This work was supported in part by grant to J.C.B.  from the 
Natural Sciences and Engineering Research Council of Canada.

\bibliography{rae_kyle_bib}

\end{document}